\documentclass{aastex631}

\makeatletter
\newcommand*{\rom}[1]{\expandafter\@slowromancap\romannumeral #1@}
\makeatother

\usepackage{longtable}
\begin{document}

\title{ON THE LIGHT-CURVEs OF DISK AND BULGE NOVAE}

\author[0009-0001-5114-4392]{Asaf Cohen}
\affiliation{Department of Physics, Ariel University, Ariel, Israel}
\author[0000-0002-7349-1109]{Dafne Guetta}
\affiliation{Department of Physics, Ariel University, Ariel, Israel}
\author[0000-0002-0023-0485]{Yael Hillman}
\affiliation{Department of Physics, Azrieli College of Engineering Jerusalem, Israel}
\author {Massimo Della Valle }
\affiliation{INAF-Capodimonte, Napoli}
\affiliation{Department of Physics, Ariel University, Ariel, Israel}
\author {Luca Izzo}
\affiliation{INAF-Capodimonte, Napoli}
\author {Volker Perdelwitz}
\affiliation{Department of Earth and Planetary Science, Weizmann Institute of Science, Rehovot,  Israel}
\author {Mario Livio}
\affiliation{ Department of Physics and Astronomy, University of Nevada, Las Vegas, 4505 South Maryland Parkway, Las Vegas NV 89154, USA }

\begin{abstract}
We examine the light curves of a sample of novae, classifying them into single-peaked and multiple-peaked morphologies. Using accurate distances from \texttt{Gaia}, we determine the spatial distribution of these novae by computing their heights, $Z$, above the Galactic plane. We show that novae exhibiting a single peak in their light curves tend to concentrate near the Galactic plane, while those displaying multiple peaks are more homogeneously distributed, reaching heights up to 1000 pc above the plane. A KS test rejects the null hypothesis that the two distributions originate from the same population at a significance level corresponding to $4.2\sigma$.

\end{abstract}

\keywords{novae, light curves, SIMBAD, automated pipeline,cataclysmic variables}

\section{Introduction}\label{sec:introduction}
The classification of novae has been a subject of scientific investigations for many decades. Early efforts in this field can be traced back to  \cite {Zwicky1936}, followed by contributions from \cite{McLaughlin1939a,McLaughlin1939b,McLaughlin1939c,Mclaughlin1945}, who alongside \cite{Pay1957} developed a "morphological" classification system based on the rate of decline of the nova's brightness from maximum light. This approach, focused on light curve behavior, formed the foundation of nova taxonomy for several decades. More recently, \cite{Duerbeck1981} expanded this morphological framework by incorporating additional observational features. These included the realization of a sub-class of novae that have a prolonged, fluctuating peak (or "multiple peaks") in the light curve before the decline, as well as drawing attention to the formation of dust in the ejecta. This marked the shift toward a more comprehensive classification system of novae. A significant advancement in nova classification came from the application of the stellar population concept, introduced by \cite{Baade1944}, whose work demonstrated that different stellar populations, with distinct physical properties, exhibit different spatial distributions within galaxies. 
Applying this concept to novae \citep{Dellavalle1994}, observed in different environments such as the spiral arms and disks of galaxies like M33, irregular galaxies like the Magellanic Clouds, and the bulges (or halo) of elliptical galaxies, provided new insights into the physics of novae. Starting in the early 1990s, several researchers explored the idea of transitioning from a purely "morphological" classification to a more "physical" one. This approach aimed to correlate nova properties, such as decline rates \cite[] {Duerbeck1990,dellaValle1992,dellaValle1995}  and spectroscopic evolution \cite[] {Williams1992,DellaValle1998,Shafter2011,Aydi2024}  with the spatial position of the systems within their galaxy, in particular differentiating between them being in the galaxy's disk or bulge. These studies suggest the existence of systematic differences between novae in bulge-dominated galaxies (e.g., M31 and the Milky Way), where Fe II novae predominate, and those in bulgeless galaxies, such as M33 or the Magellanic Clouds, where He/N and Fe IIb novae are more common \cite[] {Shafter2012,Shafter2013}.
Very recently, an interesting paper has been reported about novae in M31
\cite{Abelson2025} at same time, similar population synthesis code suggests similar results for novae in M31 \cite{Kemp2021}.

This difference likely reflects the presence of  younger stellar populations in bulgeless systems, which produce "disk" novae,  typically associated with more massive white dwarfs (WDs), compared to the "bulge" novae associated with old or very old stellar populations typical of elliptical galaxies.

These studies have shown that fast declining novae are concentrated closer to the Galactic plane, while slower declining novae are found at higher $Z$ values (distance above the galactic plane).  This result is expected. The basic condition for the formation of massive stars in one region of a galaxy rather than another is the existence of a large abundance of gas in that region, as in our Galactic plane. All known core-collapse supernovae (SNe) to date --- single stars and in binary systems --- are located in in the vicinity of star forming regions, that are located on the galactic planes of the respective parent galaxy. This implies that binary systems in our Galactic plane should host more massive stars than the stars in binary systems in the bulge. Since a nova occurs on the surface of a WD in a binary system, it can be deduced from this that the WD and donor are more massive in systems in the Galactic plane (or disk) than in systems in the bulge. Nova models show that more massive WDs tend to produce novae with faster decline times \cite[]{Prikov1995,Yaron2005,Shara2018,Hillman2020}. 

The average accretion rate has a significant impact on the decline time, with lower accretion rates leading to shorter decline times. Since lower accretion rates are associated with low-mass donors \cite[]{Hillman2020,Hillman2021b,Hillman2022a,Hillman2022b}, it can be inferred that low-mass donors result in nova explosions with shorter luminosity decline times. The underlying reason for this phenomenon is that a more powerful eruption leads to a faster decline, and powerful eruptions stem from massive white dwarfs and extended accretion phases \cite[]{Prikov1995,Yaron2005,Hillman2020,Hillman2021b}.

This indicates a competing affect of the WD and donor mass, on the resulting decline time, although, models show the WD mass to have a stronger influence \cite[e.g.,][]{Yaron2005}. Therefore, the fact that bright novae that decline quickly are observed near the plane of the Galaxy rather than at high $Z$s is highly suggestive that the observations can be interpreted as the masses of the WDs in nova producing systems near the Galactic plane being, in general, more massive than the WDs in nova producing systems in the Galactic bulge \cite[]{DellaValle2020}. 

Spectroscopic differences also appear based on the distance from the Galactic plane, with He/N novae more common near the plane and Fe II novae more frequent at high $Z$s.
This variation in nova ejecta ionization properties  has been linked to the masses of the WDs. More massive WDs, typically located in the Galactic disk, tend to produce novae with faster decline times. In contrast, systems with less massive WDs, more common in the bulge, generate novae with slower decline times. 
The mass required to trigger a thermonuclear runaway (TNR) is anti-correlated with the WD mass --- a more massive WD will require less accreted mass ($m_{\rm acc}$) \cite[]{MacDonald1983,Kovetz1985}. This is because the higher surface density attains the conditions needed for igniting hydrogen with a lower envelope mass. 
After a nova eruption, the system will relax and resume accretion until, eventually, another TNR occurs \cite[e.g.,][]{Epelstain2007}. 
Since the $m_{\rm acc}$ is primarily dependent on the WD mass \cite[e.g.,][]{Truran1986,Prikov1995,Yaron2005}, the time between two consecutive eruptions, i.e., the recurrence period ($t_{\rm rec}$), or, accretion phase, for a given WD mass ($M_{\rm WD}$) is longer for a lower accretion rate ($\dot{M}_{\rm acc}$). This means that the time between eruptions is longer for lower accretion rates and for less massive WDs. The time between eruptions has consequences on the outcome of the nova since it allows more time for the accreted hydrogen to diffuse into the outer layers of the WD’s core. This deepens the point where the TNR occurs, thus leading to more ejected mass ($m_{\rm ej}$) and a more powerful eruption \cite[]{Prialnik1984,Prialnik1986,Iben1992,Hillman2021b}. 
This culminates in nova characteristics such as the maximum brightness, the mass loss time, enrichment of the ejecta composition, etc. Extensive modeling revealed general trends that are dictated from certain system parameters. For instance, lower accretion rates lead to more ejected mass, but also to higher fusion temperatures, a brighter peak and a faster decline. More massive WDs show trends such as, higher ejecta velocities, a brighter peak and a shorter decline time \cite[]{Prikov1995,Yaron2005}. \cite{Strope2010} have recognized several different morphological behaviors for $\sim90$ Galactic novae based on 500 days post-peak, indicating the physical parameters of the system at play. 

\cite{Poggianti2018} has reported novae with prolonged, fluctuating peaks, and
\cite{Mason2020} explained this type of behavior to be possible for low mass WDs due the combination of low surface gravity and high envelope mass requiring more time for mixing.  \cite{Dubovsky2021} and \cite{Fujii2021} reported a similar behavior for V1391 Cas. 
\cite{Hillman2022b} offered support to this scenario by demonstrating via models that this feature occurs in systems with low-mass WDs ($\lesssim0.65 M_\odot$) and low accretion rates ($\lesssim10^{-10}M_\odot\rm yr^{-1}$) which results in convective mixing during the eruption.
Following this, we may deduce, that if there are regions of a galaxy that are rich in this morphological type of light curve — prolonged, fluctuating peak — this region is rich in low mass WDs ($\lesssim0.65 M_\odot$) and that they have a low average cyclic accretion rate ($\lesssim10^{-10}M_\odot\rm yr^{-1}$).

In this paper, we examine the light curves of Galactic novae,  and we explore whether the morphological shape of a nova’s light curve at maximum brightness may be linked to its location within the Galaxy. Using distances recently determined by the data release DR3 of \texttt{Gaia} \cite[]{Schaefer2022} we estimate the height of these novae above the Galactic plane, searching for any spatial correlation between light curve morphology and Galactic distribution.

\section{The distribution of novae in our galaxy}

In the Milky Way, nova populations are generally categorized by their spatial distribution and related characteristics, providing insights into their underlying stellar populations and evolutionary histories. Observations suggest that the Milky Way hosts two primary populations of novae --- those that are primarily in the Galactic disk("disk novae"), and those that primarily reside in the Galactic bugle ("bulge novae"). Each group exhibits distinct properties described below.
\begin{enumerate}
\item  Disk novae: This group includes novae that are concentrated in the Galactic disk, often closer to the Galactic plane, where young stellar populations are found. These novae are frequently observed in the spiral arms and in the vicinity of recent star formation.
Disk novae are commonly of the He/N type, characterized by strong helium and nitrogen emission lines in their spectra. Some Fe IIb novae are also present in the disk population.
Disk novae tend to occur in systems that host more massive white dwarfs (WDs). The presence of more massive WDs is associated with shorter intervals between nova eruptions and faster decline rates. This is because massive WDs require less accreted material to trigger thermonuclear runaways (TNRs), resulting in brighter and shorter-lived nova eruptions.
Disk novae are thought to represent young stellar systems with high metallicities. This connection to young populations and massive WDs is consistent with the regions of the Galaxy where star formation continues to occur, such as the Galactic plane.
\item Bulge novae:
This group comprises novae that are primarily located in the Galactic bulge, a region dominated by older, metal-poor stellar populations. They are generally found at high $Z$ values (distances from the Galactic plane), which correlates with older stellar populations.
Bulge novae are predominantly Fe II novae, identified by the presence of iron (Fe II) emission lines in their spectra which is common in environments with older, lower-mass stars. The WDs in bulge novae are typically less massive than those in the disk. Lower-mass WDs require more accreted material to reach the conditions needed for a TNR. Cases of these novae that also experience a low accretion rate, result in prolonged eruptions that decline slower. A long decline time is typical of bulge novae. Bulge novae thus serve as tracers of older stellar populations in the Milky Way.
\end{enumerate}
 
These two primary populations of novae provide valuable insights into the age, metallicity, and mass distribution of stellar populations within the Milky Way \cite{Kemp2021}. The fact that different types of novae predominate in the disk and bulge supports the idea that nova characteristics—such as spectral type, decline time, and spatial distribution—are strongly influenced by the properties of the underlying stellar population, including white dwarf mass and the age of the binary system.

Studying these populations will contribute to the understanding of the mass distribution of white dwarfs in binary systems in different galactic regions, shedding light on stellar evolution, binary star formation, and the role of novae as indicators of galactic structure and population composition.

\section{Description of the Sample}\label{sec:telescopes}

Our data source is the catalog by \cite{Schaefer2022}, which lists 402 novae. From this sample, the author identified a "golden sample" of 74 objects for which distances to the novae could be measured with sufficient accuracy, i.e., with a small error in the parallax measurements, ($p$), $\sigma p / p < 0.30$ (as given in Table 5 of \cite{Schaefer2022}). From this subset, we collected data from the literature (including the AAVSO database) to produce lightcurves for 46 objects, which are the focus of this paper. 
We chose the 46 objects that have sufficient data points  to reproduce  the lightcurves to classify them morphologically.

\begin{figure}
{\includegraphics[width=0.5\columnwidth]{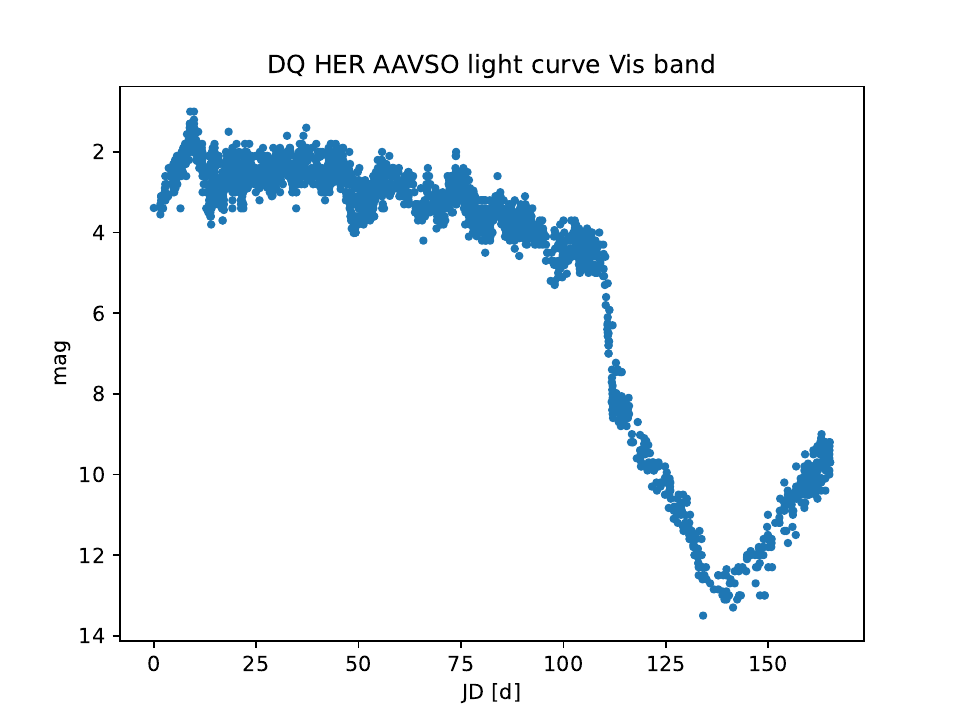}}
{\includegraphics[width=0.5\columnwidth]{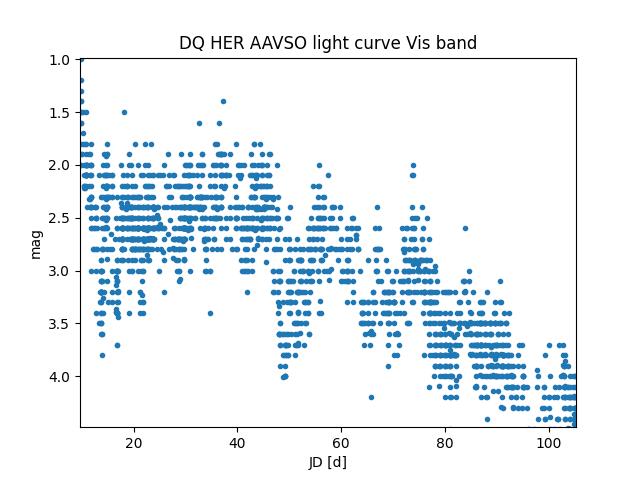}}
    \caption{Light curve of nova DQ Her as an example of a nova with a prolonged, fluctuating peak. The complete light curve (left); and a close up of the fluctuating peaks (right).}
    \label{fig:DQHER}
\end{figure}

\begin{figure}
{\includegraphics[width=0.5\columnwidth]{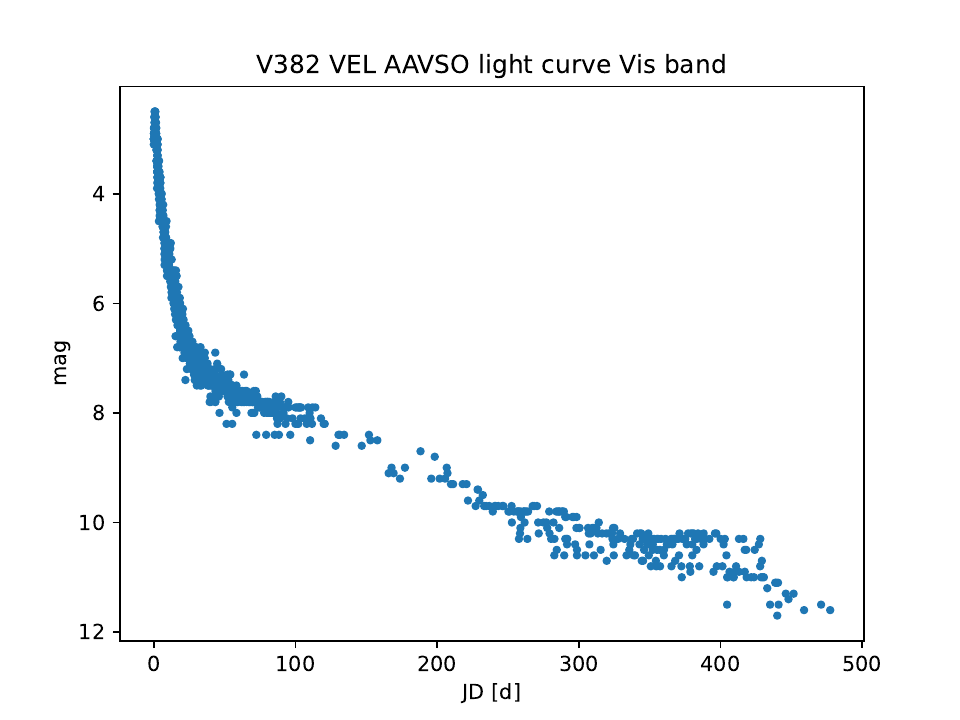}}
{\includegraphics[width=0.5\columnwidth]{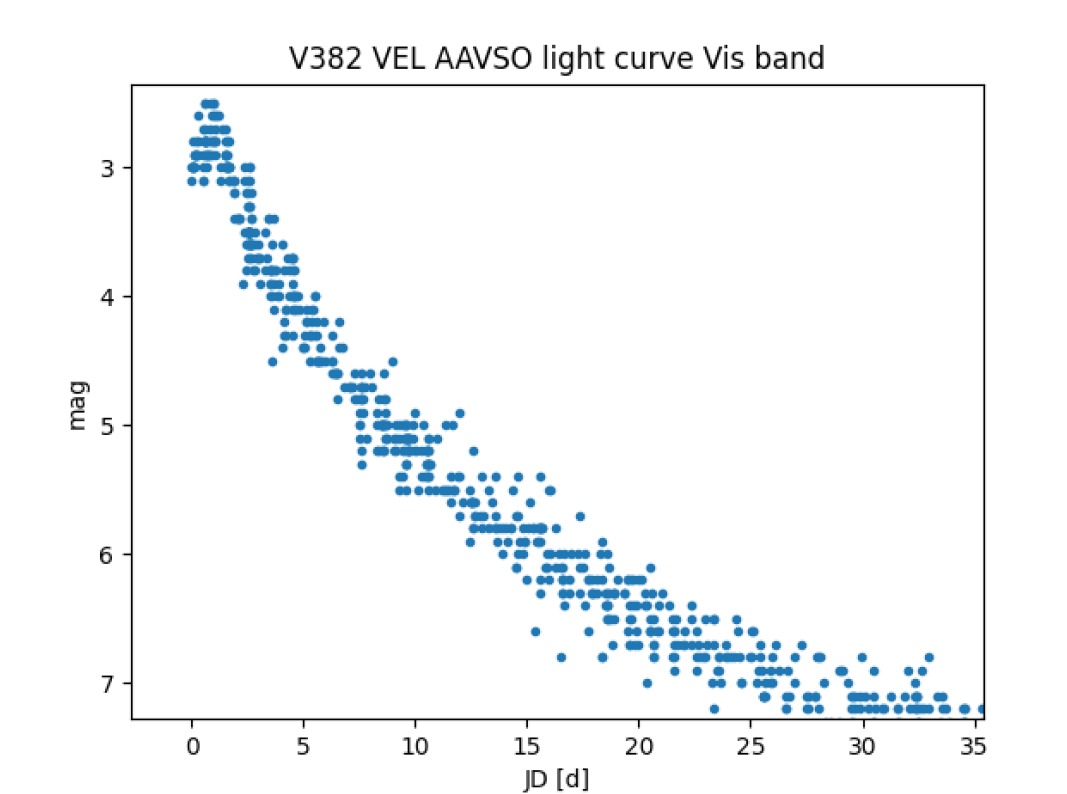}}
\caption{Light curve of nova V382 Vel as an example of a nova with one sharp peak. The complete light curve (left); and a close up of the peak (right).}
    \label{fig:V382}
\end{figure}

The analysis has been carried out following these steps:
\begin{enumerate}
\item 
We have retrieved the optical light curves for the 74 Galactic novae that have good determined distances. The primary database that has been used for this step is the AAVSO archive.
\item We have examined the morphology and timescale of each light curve in detail in order to divide the LCs into two classes: (1) novae that experienced a prolonged fluctuating peak as showed in Figure \ref{fig:DQHER} and (2) novae that exhibited one clear peak as showed in Figure \ref{fig:V382} .

\item The latitude and \texttt{Gaia} distance ($b$ and $d$ respectively) that are given in \cite{Schaefer2022}, have been used to calculate the altitude above or below the Galactic plane: $Z=d{\rm sin} (b)$.  
\item Using the derived $Z$ of each nova in both classes, we have constructed a distribution in order to search for a correlation.
\item Additionally, we have extracted for all the novae in both classes, the time to decline from the peak by  three magnitudes, $t_3$. For convenience, we express the decline time  as $V_d$, which is  the number of magnitudes, 3, divided by  $t_3$. In order to estimate  $t_3$, we have fitted the LC with different functions and found how long it took the LC to decline by 3 magnitudes from the first peak.

\end{enumerate}

\section{Results}

\begin{figure}[h!]
   \centering
    \includegraphics[scale=0.7]{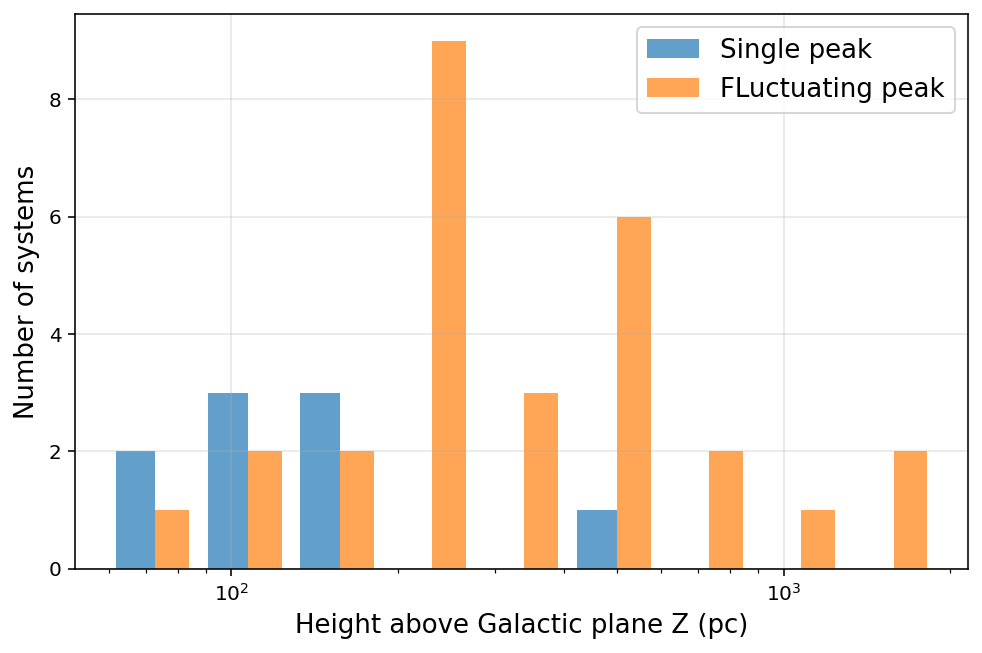}
   \caption{Distribution of the distance from the Galactic plane ($Z$)  for both nova classes.}
    \label{fig:energy_levels}
\end{figure}

Of the 74 novae with well determined \texttt{Gaia} distances, 46 had attainable data with sufficient data points to construct a light curve. Of these, 18 novae are characterized by a distinct peak that decreases by three magnitude in less than three days. These novae constitute the subclass of "single peak novae". The remaining 28 novae exhibit light curves that are characterized by a prolonged, fluctuating peak that lasted a few tens of days. These novae constitute the subclass of "fluctuating peak novae" (or "multiple peaks"). The distance from the Galactic plane ($Z$) was calculated for all 46 of the novae and a distribution based on $Z$ was produced for each subclass. The geographical data for all 46 novae, including the calculated $Z$, are detailed in Table \ref{long1}. Figure \ref{fig:energy_levels} shows that the LCs with one peak tend to be concentrated towards smaller $Z$s, while the multiple peak novae extend all the way up to about $Z\sim$1000 pc.  The physical basis for apportioning the novae in Figure 3 is the size of the Galactic plane, which is approximately $<z>~\sim 100$ pc. We have plotted Figure 3 using a logarithmic scale, which is more suitable for highlighting the differences between the two distributions, especially when dealing with a large range of values: from 100 pc to 1000 pc.  We carried out a KS test to assess the probability that the two distributions were drawn from the same population and found the probability of $P\sim9\times 10^{-5}$ allowing us to reject this possibility at approximately the $3.7\sigma$ confidence level. A Monte Carlo simulation (100,000 simulated distributions of $Z$) was also performed, yielding a probability of reproducing the observed distribution by chance of $P\sim 2\times 10^{-5}$, corresponding to approximately 4.2$\sigma$. 
A comment regarding potential selection effects that could shape the observed distribution seems in order. While it is possible that some multiple-peak novae, which are generally intrinsically fainter than single-peak novae, may be missed if they occur in the crowded Galactic plane, there is no selection effect that can explain the lack of bright novae at higher heights above the Galactic plane. In other words, even though faster novae are brighter and more easily detected in the plane, there’s nothing preventing their detection outside of it.  Moreover, even if slower novae might be missed in the plane, this doesn’t change the fact that multiple-peaked novae are spread over a much broader range of heights, extending up to $\sim 1000$ pc. Therefore, the non-uniform distribution we observe cannot be attributed to a selection effect.

This result strengthens the outcome of the KS test. This result serves as a starting point for the next stages of this research, by raising the question of why single peaked novae favor dense environments like the Galactic disk, while the fluctuating peaked novae are randomly distributed. This is likely related to the mass of the WD and the rate at which it accretes mass, as suggested in previous works \cite[]{Poggianti2018,Mason2020,Hillman2022b}. However, a detailed investigation and modeling are required in order to fully understand the underlying causes, which constitutes one of the primary goals of this proposed research.  

\newpage


\begin{longtable}[c]{| c | c | c | c | c | c | c | c  | c |}
  \caption{48 confirmed infrared light curves of novae. All details are according to the parameters listed in the catalogue \cite[]{Schaefer2022}. Distance from latitude ($Z$) was calculated by $Z=d\sin (b)$; $t_3$ is the time to decline by three magnitudes; $V_d$ is $3/t_3$; and the classification by peak type is "op" (one peak) or "mp" (multiple peaks). }
   \label{long1}\\
 \hline
 \multicolumn{7}{| c |}{Comparison novae}\\
 \hline
  \textit{Nova} & \textit{d(\rm\texttt{Gaia})} & \textit \textit{$b$(latitude)} & {$Z$(distance from latitude)} & {$t_3$}  & {$V_d$} & {classification}  \\
  \textit{name} &  [$pc$] & [$pc$] & [$pc$] & [days]  &[$\rm 1/days$]  & {by peak type}\\
 \hline
 \endfirsthead

 \hline
 \multicolumn{7}{|c|}{Comparison Novae}\\
 \hline
 \textit{Nova} & \textit{d(\rm\texttt{Gaia})} & \textit \textit{$b$(latitude)} & {$Z$(distance from latitude)} & {$t_3$}  & {$V_d$} & {classification}  \\
  \textit{name} &  [$pc$] & [$pc$] & [$pc$] & [days]  &[$\rm 1/days$]  & {by peak type}\\
 \hline
 \endhead

 \hline
 \endfoot

 \hline
 \multicolumn{7}{| c |}{End of Table}
 \\
 \hline
 
 \hline
 \endlastfoot

V603 Aql   &  314  &  0.829   &   4.543  &   15  & 0.2 & op \\ 
DQ Her   &  501  &   26.44   & 223.1    &   90  & 0.033 & mp\\  
RR Pic  & 511  &   -25.672  &  221.4  &   130 &0.023 & mp  \\  
CP Pup  & 814 &   -0.835  &   11.86   &  12   &  0.25 & op \\ 
V841 Oph    &  823  &  17.8    &  251.6  &   90    & 0.033 & mp\\ 
GK Per    & 442  &  -10.104     &  77.54    &  11   &0.272 & op \\ 
HR Del    &  958  &  -14   &  231.8    &  220   & 0.013 & mp\\
V533 Her    &  1202  &  24.273     &   494.1   & 75   & 0.04& mp\\
T CrB   &   910  &   48.2  &    678.4   &   8  & 0.375& mp\\ 
CP Lac   &  1170  &   -0.837   &  17.09   &   10   &  0.3 & op\\
 V1016 Sgr     &  2670  &  -4.8  &  223.4  &  77   & 0.039& mp\\ 
RS Oph &  2710  &  10.4    &   489.2    &   10   & 0.3& op\\
V598 Pup    & 1840  &  -13.8    &438.9   &  220  & 0.013 & mp  \\
FH Ser      &  1060  &   5.786   &   106.9   &  60   & 0.05 &mp \\
V382 Vel     & 1530  &   5.8   &  154.6   &  12   & 0.25&op \\
V5558 Sgr  &  8140  &   -6.5   &   921.5   &  140   &0.021& mp \\
V842 Cen   &  1379   &   2.452   &   59  & 40 & 0.075& mp \\
DN Gem    & 1365   &    14.71  &  346.6    &  22   & 0.136& mp \\
 V721 Sco  &  1860  &   -2.4  &   77.89   &    17    & 0.176&op \\
V1280 Sco    &  3500  &   6.6   &   402.3   & 14 &  0.214& mp \\
V1974 Cyg   & 1631  & 7.819 & 221.9 & 215  & 0.013 & mp\\
NQ Vul  &  1080  &    1.29    &  24.31     &    2    & 1.5& op \\
T Pyx  &   3185 &    9.707    &   537     &    55   & 0.054& mp \\
V446 Her   &  1361 &   4.708    &   111.7   &   35     & 0.085& op \\
KT Eri   &  1260  &     4.7    &  103.2     &     40    &  0.075&op \\
CI Aql    &  2760  &   -0.8   &     38.54   &      30    & 0.1& op \\
YZ Ret  &  2390 &     -46.4   & 1731 & 17 &  0.176 & mp \\
V705 Cas  &  2157 & -4.096 &  154.1  &  52   &   0.057& mp  \\
V888 Cen   &   2940    &   2.6 &  133.4 & 58 & 0.051 & mp \\
V1494 Aql  &  1239  & -4.742 & 102.4 & 15 & 0.2 & op \\
V368 Aql   &  2450    &  -2.6  & 111.1  &  & & mp \\
V476 Cyg  &  665   &    12.42 &    143 & 13 &  0.230 & op  \\
GI Mon   &  2849   &    4.749& 235.9 &   110   & 0.027 & mp\\
V4065 Sgr  &   7860   & -4.5 &  616.7  &  & &mp \\
DK Lac   &   2517   &   -5.352 &  234.8    & 50   & 0.06 &mp\\
PW Vul   &    2420    &  5.197 &    219.2  &  70 &0.042& mp\\
 V1500 Cyg  &   1287   &  -0.073  &  1.64  &  5 & 0.6 & op \\
CT Ser  &   2550    &  47.6   & 1883    &   130   &  0.023 &mp \\
V392 Per &    3400    &    0.9    &   53.4    &   10   &  0.3&op \\
 V1369 Cen  &   640  &  2.7 &   30.15 & 10   & 0.3 & op \\
V1330 Cyg &    2880   &   -5.5  & 276 & 200 & 0.015 &mp\\
V2467 Cyg   &   1840  &  1.8  &   57.8   & 13 &  0.230 &op  \\
V1535 Sco  &  7790  &  3.9   &  529.8   &   26   &  0.115&mp \\
V5667 Sgr  &   7690   &  -4  & 536.4  &  160 & 0.018 & mp\\
V450 Cyg  &   4590 &  -6.5  &  519.6  & 200 & 0.015 &mp  \\
QU Vul  &    1740  & -6   & 181.9 & 36 &  0.083 & op \\

 \end{longtable}

\noindent



\section{Discussion}

The aim of this study was to investigate the Galactic distribution of nova light curves, classified into two groups according to their behavior at maximum light: novae with a sharp, single peak and fast decline, and novae with multiple peaks and prolonged fluctuations and slower photometric evolution. According to the concept of “stellar population” introduced by \cite{Baade1944}, by identifying regions of the Galaxy that are rich or poor in one of these LC types directly implies a range of properties of the parent stellar population and system parameters characterizing the physics of nova explosions.

We have shown that the distribution of light curves with a single peak versus a fluctuating peak, as a function of height above the Galactic disc, $Z$, indicates that single-peak LCs tend to be concentrated at small $Z$s, while multiple-peak LCs extend up to about $Z\sim$1000 pc. The probability of reproducing the observed distribution by chance was $P\sim 2\times 10^{-5}$ corresponding to approximately 4.2$\sigma$. This result has a rather straightforward interpretation: intrinsically bright novae produced by systems with massive WDs $\sim 1.2M_\odot$ are the end products of relatively massive progenitor stars, between 6 and 8 $M_\odot$ \cite[]{Cumming1994,Kemp2021}  These WDs tend to be located predominantly in the Galactic plane, a region rich in gas and ongoing star formation. For such a WD to reach the conditions required for a nova eruption, it needs to accrete a relatively small amount of material ($m_{\rm acc}~\sim10^{-6}-10^{-5}M_\odot$ for a $\sim1.2M_\odot$ WD). This leads to the ejection of a lighter envelope,  in a single, sharp episode, producing the smooth, single-peak light curves typical of these systems.

In contrast, intrinsically faint nova systems characterized by low mass WDs ($\sim0.65M_\odot$) and slower evolution, can be found at greater heights from the Galactic plane, spanning from the thin to the thick disc. Since these WDs are less massive, they are able to hold more accreted mass before reaching the conditions required for igniting the hydrogen and initiating the nova eruption, hence they accrete more massive envelopes, of about 10 times more mass than those of $\sim1.2M_\odot$ WDs. 

For all cases, after igniting the TNR, the amount of mass that will be ejected ($m_{\rm ej}$) is strongly dependent on the average cyclic accretion rate \cite[e.g.,][]{Yaron2005,Hillman2019}. For high rates, only a small fraction of the accreted mass ($m_{\rm acc}$) will be ejected, while for low rates the entire accreted envelope may be ejected (and often even $m_{\rm ej}>m_{\rm acc}$). For high WD masses, the ejection will occur quickly regardless of the average cyclic accretion rate that preceded the eruption, because even the entire envelope is still a small amount of mass. But for low mass WDs, which have relatively low surface gravity, and a relatively high mass to be ejected, the process may be slow, with mixing occurring in the deeper layers of the envelope during the ejection, resulting in fluctuations of the light curve \cite[]{Mason2020,Hillman2022b}.

Considering the energy budget of a typical nova explosion, approximately $10^{45}$ erg \cite[]{Gallagher1978}, a single episode is sufficient to expel the lighter envelopes at velocities as high as ~$3000-4000$ km/s in novae with massive WDs, thus producing a sharp, single peak maximum. On the other hand, although novae with lighter WDs accrete envelopes roughly 10 times more massive, the lower ejection velocities, typically as low as few hundred km/s, imply that each ejection event only uses a fraction of the total energy budget. Consequently, a substantial amount of energy remains available to fuel subsequent ejection episodes, resulting in the multiple peaks observed in these light curves.

\section{Conclusions}

Our study investigates the morphological classification of nova light curves and their spatial distribution within the Galaxy, by using the latest \texttt{Gaia}-derived distances. By examining 46 Galactic novae, we identify two distinct groups: single-peaked light curves, which decline rapidly, and multiple-peaked light curves, exhibiting a more complex behavior. Our analysis reveals a significant correlation between light curve morphology and height above the Galactic plane. Specifically, single-peaked novae are predominantly found closer to the Galactic plane, while multiple-peaked novae are distributed across a wider range of altitudes, extending all the way to $\sim  1000$ pc.

This distribution pattern supports the hypothesis that the physical properties of novae, specifically the WD mass, varies with Galactic location. Our findings are consistent with previous ideas suggesting that more massive WDs, which produce faster declining novae, are more likely to be found in the gas-rich Galactic disk. Nova systems with lower mass WDs and slower decline rates are more prevalent at higher altitudes, reflecting older stellar populations.  This idea of two distinct groups is not in conflict, but rather aligns with other findings such the recent result by \cite{Abelson2025}, who showed that the nova population in M31 can be described by a delay-time distribution characterized by two separate peaks, further supporting the notion that novae exhibit diverse physical and temporal properties depending on their environment. 

In this context, the classification into two distinct groups, based on physical features, provides a more comprehensive and simplified framework for understanding nova behavior, moving beyond the purely morphological classification that has dominated nova studies for decades. The results presented here reinforce the idea that the morphological and physical characteristics of novae are intimately connected to their Galactic environment. Future work could further refine these correlations by incorporating a larger dataset and exploring additional parameters that influence nova behavior.

\bibliography{biblio.bib}{}
\bibliographystyle{aasjournal}

\end{document}